\documentclass[aps,a4paper,10pt,twocolumn,footinbib]{revtex4}
\usepackage{amsmath}
\usepackage{amsfonts}
\usepackage{amsfonts}
\usepackage{mathrsfs}
\usepackage[mathscr]{euscript}
\usepackage[dvips]{graphicx}
\usepackage{fancyhdr}
\usepackage{colordvi}
\usepackage[hypertex=true]{hyperref}
\usepackage{epsfig}
\usepackage{color}

\begin{document}
\title{On active drains and causality}
\author{Paul Kinsler}
\email{Dr.Paul.Kinsler@physics.org}
\affiliation{
  Blackett Laboratory, Imperial College,
  Prince Consort Road,
  London SW7 2AZ, 
  United Kingdom.
}

\lhead{\includegraphics[height=5mm,angle=0]{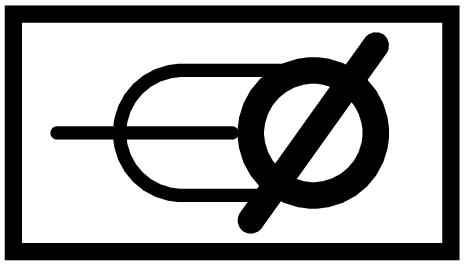}~~LFIADC}
\chead{~}%On active drains and causality}
\rhead{
\href{mailto:Dr.Paul.Kinsler@physics.org}{Dr.Paul.Kinsler@physics.org}\\
\href{http://www.kinsler.org/physics/}{http://www.kinsler.org/physics/}
}
%\lfoot{\thesection . \thesubsection; ~~~~ (\yymmdddate\today:\currenttime) }
%\rfoot{{\large \emph{Not for redistribution}}}

\begin{abstract}

The concept of an \emph{active drain}
 has been used recently to provide an optical element 
 which manages to perfectly sink 
 incoming electromagnetic radiation.
Here I show that without prior knowledge of the incoming signal,
 an element attempting respond as  
 an active localized drain cannot succeed.

\end{abstract}

\date{\today}
\maketitle
\thispagestyle{fancy}

%
% =======================================================================
%\section{Introduction}\label{S-intro}

There is a current interest in
 using time-reversed sources as a kind of perfect sink 
 (or ``drain'') 
 {designed to exactly cancel incoming signals} 
 \cite{deRosny-F-2002prl,Chong-GCS-2010prl,Leonhardt-2009njp}.
These are active, 
 not passive devices, 
 and need to be perfectly matched to the incoming radiation -- 
 i.e. both spatially \emph{and} temporally matched.
Here I {leave the spatial properties to one side}
 and instead focus {solely} on the temporal behaviour.
I ask the question:
 \emph{If the active drain does not know what signal is about to arrive, 
 can it exactly cancel the incoming field regardless?}

{Thus here I consider a more general case than was considered 
 in the three cases mentioned above 
 (i.e. \cite{deRosny-F-2002prl,Chong-GCS-2010prl,Leonhardt-2009njp}), 
 where knowledge of the incoming radiation is known in advance
 and so built into the device.
This can be seen from how De Rosny et al.'s acoustic sink
 is explicitly constructed from a time-reversal
 of the known source, 
 and how Chong et al.'s time-reversed laser
 \cite{Chong-GCS-2010prl} demands }
``specific conditions of coherent monochromatic illumination''.
For the mirrored version of Maxwell's fisheye lens
 \cite{Leonhardt-2009njp},
 whilst its claimed perfect imaging capabilities
 are a matter of ongoing debate 
 \cite{Blaikie-2010njp-fisheyeC,Leonhardt-2010njp-fisheyeR,
       Kinsler-F-2010njp-fisheye,Sun-H-2010arxiv,Merlin-2010arxiv},
 nevertheless all parties seem to agree
 that \emph{if} it can do so\footnote{{And whether or not
  such a process still constitutes
  what might normally be described as ``imaging''.}}, 
 it relies on the presence of an active drain.

%
% =======================================================================
\section{The response of an active drain}\label{S-response}

In {a more general} environment
 where the properties of the incoming signal properties
 are not predetermined and already designed into the active drain, 
 the drain element will have to follow some causal response driven
 by the source $S(t)$.
{In the case of electromagnetism, 
 the drain response might in essence be just some more complicated version
 of an ordinary dielectric polarization, 
 which might e.g. follow a Lorentz or Drude form \cite{RMC}.
I therefore start by 
 writing down a general differential equation
 for the response of the drain element $P(t)$
 using a summed series of time derivatives 
 (where $\partial_t \equiv d/dt$).}
{This then allows as wide as possible 
 a range of responses, 
 and is unlikely to exclude behaviours that might
 turn out to be useful\footnote{{As pointed out
  by the second reviewer,
  in addition to allowing derivatives of the source term, 
  it is also legitimate to consider further time-delayed 
  properties of the source.}}.
This general response is
~
{\begin{align}
  \sum_{n=0}^{N}
  T_n %\left(T_n\right)^n
  \partial_t^n P(t)
&=
  \sum_{m=0}^{N-1}
  a_m
  \partial_t^m
  S(t-\tau) %e^{\imath \phi}
.
\label{eqn-t-domain}
\end{align}}
Here the $T_n$ {control the dynamics of the drain element}, 
 while the
 {$a_m$} allow for the coupling
 between the drain element and the {properties of the}
 incident signal field $S$\footnote{{E.g. for a Lorentz response, 
  we would have $T_2=1$, $T_1=\gamma$ (loss), $T_0=\omega_0^2$, 
  {and only $a_0 \neq 0$}; 
 the signal S(t) would be replaced by the driving from the local value
 of the electric field.}}.
The drain element experiences a signal field delayed by $\tau$
 from that produced by the source according
 to the path length\footnote{E.g. 
  in the mirrored fisheye lens of \cite{Leonhardt-2009njp}, 
  all optical path lengths from source to drain are the same, 
  and correspond to a phase shift of $\phi = \omega \tau = 20\frac{1}{4}\pi$.}.
{In the description here I do not consider how propagation might affect
 the source signal before it reaches the location of the drain.
Here,
 $S$ defines what that signal has become when it reaches the drain, 
 although I leave in the time-delay $\tau$ to emphasize the retardation
 between the generation of the signal and its reception.} 
Note that we cannot {have an $n=0$ term by itself}, 
 because contributions like ``$T_0 P= a_0 S$''
 specify an identity, 
 not a causal relationship{;
 more generally the derivatives on the LHS should be 
 at least one order higher than those on the right.}
In the frequency domain, 
 where time derivatives %(i.e. $\partial_t$) 
 convert to factors of $- \imath \omega$, 
 eqn. \eqref{eqn-t-domain} becomes
~
\begin{align}
  \sum_{n=0}^{N}
  \left( - \imath \omega \right)^n T_n
  P(\omega)
&=
  {\sum_{m=0}^{N-1}
  a_m}
  S(\omega) e^{\imath \omega \tau}
.
\label{eqn-w-domain}
\end{align}
This can then be rearranged into 
$  P(\omega)
=
  f(\omega)
  ~
  S(\omega)
  e^{\imath \omega \tau}
$ for a response function $f$ which is 
 is a rational function 
 of frequency $\omega$, i.e.
~
\begin{align}
  f(\omega)
&=
  \frac{{\sum_{m=0}^{N-1} \left( - \imath \omega \right)^m a_m}} 
       {\sum_{n=0}^{N}   \left( - \imath \omega \right)^n T_n } 
\label{eqn-w-domain-rhs}
.
\end{align}
For a finite maximum $N$, 
 appropriate choices of $T_n$ and $a_m$ should ensure
 that $f$ has sufficiently simple poles 
 below some line parallel to the real $\omega$ axis \cite{Nistad-S-2008pre}.
{Further, 
 a non-zero $T_N$ and
 the restriction of the $m$ summation to a maximum of $N-1$
 guarantees that $f$ 
 has the appropriate limiting behaviour of at least $\sim \omega^{-1}$}
 as $\omega \rightarrow \infty$.
Thus 
 $P$ can be shown to remain causal in the sense that 
 it satisfies the Kramers Kronig relations \cite{LandauLifshitz}.

However, 
 in addition to $P$'s response to the driving from $S$, 
 it also needs to behave like a drain:
 i.e.
 we need that $P(t) = -S(t-\tau)$ % e^{\imath \omega \tau}$ 
 or $P(\omega) %= - S^*(\omega) e^{\imath \omega \tau}
  = - S(\omega) e^{\imath \omega \tau}$.
Without this perfect correspondence, 
 the active drain will not sink the incoming signals
 from the source.
{Note that the cancellation of fields by a drain is not ``loss''
 in the sense of some irreversible dissipation, 
 but is instead a carefully arranged \emph{destructive interference}.}
Hence, 
 for the drain to work, 
 we need
~
\begin{align}
  \sum_{n=0}^{N} \left( - \imath \omega \right)^n T_n 
&=
 -
  {\sum_{m=0}^{N-1} \left( - \imath \omega \right)^m a_m}
,
\label{eqn-w-domain-condition}
\end{align}
 which amounts to demanding that 
 {two polynomials of \emph{different order} are somehow equal}.
{Crucially,}
 since the  $T_n$ {and $a_m$
 are fixed parameters of the response function, 
 the equality will not hold over some finite frequency interval, 
 but only at specific intersection frequencies.}

{Now consider an electromagnetic drain whose behaviour is expressed
 in terms of dielectric polarization $P$, 
 with a signal arriving through a dispersionless linear medium
 as an electric field $E$, 
 and impinging from all directions on the drain
 (such as the image point of a Maxwell's fisheye lens \cite{Luneberg-MTO}
 as discussed by Leonhardt \cite{Leonhardt-2009njp}).
Here we need the drain polarization $P(t)=-\epsilon E(t)$, 
 so that the displacement field $D=\epsilon E + P$ vanishes. % {--
% just as if there were no signal field at all (i.e. $E=0$)}.
This follows because 
 for situations without magnetic sources (as considered here), 
 the second order wave equation can be written as
 $\nabla^2 E = - \mu \partial_t^2 D$.
Thus if $D=0$ is always true at the drain position,
 then by simple integration, %in Cartesian coordinates,
 $E$ at the drain position can at most have
 a linear spatial variation offset by some constant background.
For a symmetric situation
 both linear and background terms vanish --
 the linear is incompatible with symmetry, 
 and the offset must remain at the value 
 established before the signal field
 arrived at the drain (i.e. zero).
In simple terms, 
 without the drain's polarization $P$, 
 an incoming signal would pass through the drain point 
 and become an outgoing wave; 
 but the drain polarization creates a wave anti-phase to the outgoing signal, 
 causing perfect cancellation:
 only the incoming signal field survives.
In asymmetric cases, 
 such as where a field impinges onto (e.g.) the left hand side 
 of a drain designed to cancel only the outward-going fields on the right, 
 $D=-\epsilon E$
 is then the appropriate criterion\footnote{This is because
 the drain polarization must radiate in both directions, 
 but only the right hand part cancels the outward signal field.}.
This can be shown using e.g. 
 directional decompositions of the wave equation
 \cite{Kinsler-2010pra-fchhg,Kinsler-2010pra-dblnlGpm}.}

{Note that here I have chosen a relatively easy task,
 since I consider drains that only match the temporal properties
 of a signal at a single point.
In contrast, 
 drains that aim to e.g. 
 cancel a signal over some spatial region,
 or minimise reflections and transmission,
 would suffer more constraints and so be either 
 harder to implement or suffer worse performance.}

%
% =======================================================================
\section{An example}\label{S-example}

\begin{figure}
\includegraphics[angle=-0,width=0.80\columnwidth]{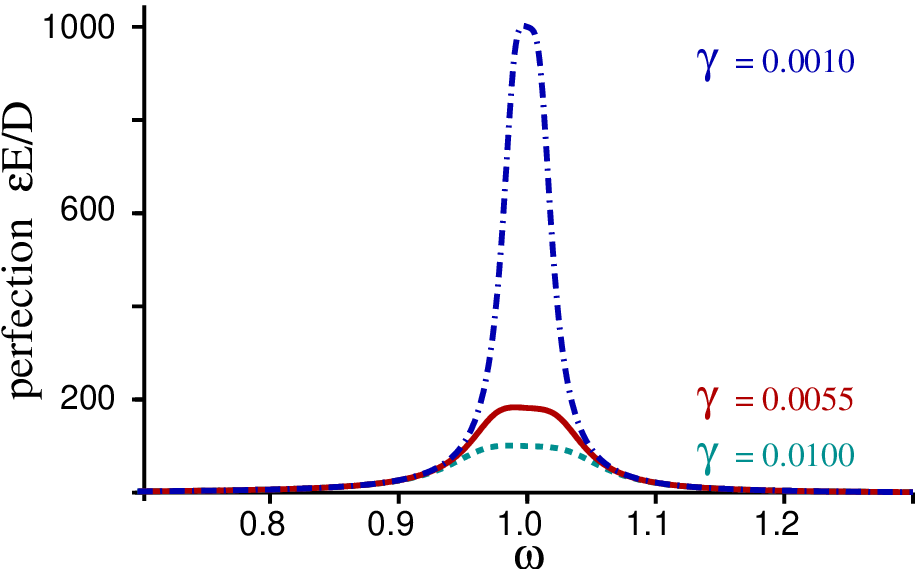}
\caption{
The spectral ``perfection'' $|\epsilon E / D|$ of the active drain model
 as calculated from eqn. \eqref{eqn-quarticmodel-DoE}, 
 for $T_4=1$, $T_2=2$, and three values of loss $\gamma$.
A value over 100 implies signal cancellation to within 1\%.
}
\label{fig-01-perfecto}
\end{figure}

Consider a device designed to act like a point-like active drain
 with a frequency independent response near its operating frequency $\omega_x$.
This behaviour can be satisfied perhaps most simply 
 by requiring a response function quartic in time derivatives.
To make the example more concrete, 
 assume this is an electromagnetic problem 
 in a medium with a linear background permittivity of $\epsilon$.
Therefore
 for incident signal (electric) field $E$
 and local response (dielectric polarization) field $P$, 
 we have 
 $T_4 \partial_t^4 P + T_2 \partial_t^2 P + \gamma \partial_t P 
   = \epsilon E$, 
 where I have used $\gamma$ instead of $T_1$ to indicate the loss
 time-scale of the device response.
In the frequency domain, 
 the response is
~
\begin{align}
  \left[
    T_4 \omega^4 
   -
    T_2 \omega^2
   -
    \imath \gamma \omega
  \right]
  P
&=
 \epsilon
 E
.
\label{eqn-quarticmodel-w}
\end{align}
A ``perfect drain'' condition occurs when these two fields ($E$ and $P$)
 exactly cancel (i.e. $P = -\epsilon E$), 
 giving a null displacement field $D = \epsilon E + P = 0$.
Thus 
~
\begin{align}
  \frac{D}{\epsilon E}
&=
  \frac{\epsilon E + P}
       {\epsilon E}
%\\
\quad
=
%  \frac{
%    \left(
%      T_4 \omega^4 - T_2 \omega^2 - \imath \gamma \omega
%    \right)
%    P
%   +
%    P
%  }
%  {
%    \left(
%      T_4 \omega^4 - T_2 \omega^2 - \imath \gamma \omega
%    \right)
%    P
%  }
%\\
%&=
  \frac{
    \left(
      T_4 \omega^4 - T_2 \omega^2 - \imath \gamma \omega
    \right)
   +
    1
  }
  {
    \left(
      T_4 \omega^4 - T_2 \omega^2 - \imath \gamma \omega
    \right)
  }
.
\label{eqn-quarticmodel-DoE}
\end{align}
This means that if 
 $T_4 \omega^4 - T_2 \omega^2 - \imath \gamma \omega + 1 \simeq 0$,
 we will be close to having constructed an active drain.
Further, 
 if the operating frequency $\omega_x$ is chosen to be
 an extremal value of $T_4 \omega^4 - T_2 \omega^2$, 
 then the frequency response will be nearly flat; 
 such an extrema occurs at $\omega^2 = T_2 / 2T_4$.
This has $T_4 \omega^4 -T_2 \omega^2 = T_2^2/4T_4$; 
 so to get the desired behaviour at $\omega_x$
 we need $T_2^2 = 4 T_4$.
The performance of the drain near $\omega_x$ will be good if 
~
\begin{align}
  \left|
      T_4 \omega^4 - T_2 \omega^2 - \imath \gamma \omega 
  \right|
\ll
  1
.
\label{eqn-eg-inequality}
\end{align}
In the vicinity of $\omega_x$,
 i.e. at $\omega = \omega_x + \delta$,
 we have that the LHS of the inequality
 in eqn. \eqref{eqn-eg-inequality} varies as
~
\begin{align}
  \left|
    T_4 \omega^4 - T_2 \omega^2 - \imath \gamma \omega 
  \right|
&\simeq
  \left|
    2 T_2 \delta^2 - \imath \gamma \omega_x
  \right|
=
%\\
%&\simeq
  \sqrt{
    4 T_2 \delta^4 + \gamma^2 \omega_x^2
  }
,
\end{align}
 so that we want both 
 $(\delta/\omega_x)^2 \ll 1 / 2 T_2 \omega_x^2$
 and
 $\gamma \omega_x \ll 1$.
Thus the device will only work for low loss 
 and over a bandwidth 
 much smaller than its designed operating frequency; 
 sample results are shown on fig. \ref{fig-01-perfecto}.
This situation is similar to that 
 regarding causal constraints on negative refractive index
 or perfect lenses \cite{Kinsler-M-2008prl,Skaar-2010arxiv-cipam}.

%
% =======================================================================
\section{Conclusion}\label{S-conclusion}

Causality restricts the perfect operation of an active drain
 to arriving signals that happen to match 
 a \emph{pre-specified} frequency of operation.
However,
 a further complication is that no realistic source is exactly
 single frequency, 
 it is only ever approximately so.
This is because 
 even if we could eliminate unwanted frequency fluctuations in the source, 
 the process of switching it on (or off) 
 necessarily involves additional frequency components.
Thus, 
 since no more than a single frequency component of the signal 
 could ever be canceled out,  
 an active drain will \emph{never} be perfect --
 except, 
 of course, 
 if causality is side-stepped by 
 agreeing in advance on the signal and its timings
 (as in \cite{deRosny-F-2002prl,Chong-GCS-2010prl,Leonhardt-2009njp}).

The example given shows that an approximate active drain 
 can be achieved over a finite (albeit small) bandwidth, 
 which in well controlled experimental situations
 should suffice to demonstrate some basic principles.
However, 
 outside carefully pre-arranged or restricted circumstances,
 an active drain cannot be guaranteed to work.

%\begin{acknowledgments}
\noindent
\emph{Acknowledgment:} 
 support from EPSRC (EP/E031463/1)
%\end{acknowledgments}

%
% =======================================================================
%\bibliography{/home/physics/_work/bibtex}

%
% =======================================================================

\end{document}